\documentclass[preprint,12pt]{elsarticle}
\usepackage{amssymb}
\journal{Physics Letters B}
\begin{document}
\begin{frontmatter}
\title{\bf Matter stability in modified teleparallel gravity }
\author{A. Behboodi}%
\ead{a.behboodi@stu.umz.ac.ir}

\author{S. Akhshabi}%
\ead{s.akhshabi@umz.ac.ir}

\author{K. Nozari}%
\ead{knozari@umz.ac.ir}

\address{Department of Physics, Faculty of Basic Sciences, University
of Mazandaran, P. O. Box 47416-95447, Babolsar, IRAN} \vspace{1cm}

\begin{abstract}
We study the matter stability in modified teleparallel gravity or
$f(T)$ theories. We show that there is no Dolgov-Kawasaki
instability in these types of modified teleparallel gravity
theories. This gives the $f(T)$ theories a great advantage over
their $f(R)$ counterparts because from the stability point of view
there isn't any limit on the form of functions that can be chosen.\\
\begin{description}
\item[PACS numbers]
04.50.Kd
\item[Key Words]
Teleparallel Gravity, Matter Stability
\end{description}
\end{abstract}
\vspace{2cm}

\end{frontmatter}
\section{Introduction}
It was Einstein who soon after formulating his theory of general
relativity, first introduced the idea of teleparallel gravity
\cite{Ein30}. In this new theory, a set of four tetrad (or vierbein)
fields form the orthogonal bases for the tangent space at each point
of spacetime and torsion instead of curvature describes
gravitational interactions. Tetrads are the dynamical variables and
play a similar role to the metric tensor field in general
relativity. Teleparallel gravity also uses the curvature-free
Weitzenbock connection instead of Levi-Civita connenction of general
relativity to define covariant derivatives \cite{Weitz23}.

After its first introduction, further important developments were
made by several pioneering works in teleparallel gravity and it has
been shown that teleparallel Lagrangian density only differs with
Ricci scalar by a total divergence \cite{Haya79,Ald10}. This shows
that general relativity and teleparallel gravity are dynamically
equivalent theories where the difference arises only in boundary
terms. However there are some fundamental conceptual differences
between teleparallel theory and general relativity. According to
general relativity, gravity curves the spacetime and shapes the
geometry. In teleparallel theory however torsion does not shape the
geometry but instead acts as a force. This means that there are no
geodesics equations in teleparallel gravity but there are force
equations much like the Lorentz force in electrodynamics.

Recently with the discovery of accelerated cosmic
expansion\cite{SN98}, modifying gravity beyond general relativity
has generated much interest. One way to modify gravity is to replace
the GR Lagrangian density, $R$, with a general function of Ricci
scalar. This approach leads to the so called $f(R)$ theories of
gravity \cite{FR}. Similarly one can try to modify gravity in the
context of teleparallel formalism and replace the teleparallel
Lagrangian density, $T$ with a general function of $T$ which leads
to the generalized teleparallel gravity or $f(T)$ theories
\cite{FT}. The resulting field equations in $f(T)$ theories are
second order equations and are much simpler than the fourth order
equations that appear in metric formalism of $f(R$) gravities.

It has been shown that $f(T)$ theories can explain the present time
cosmic acceleration without resorting to some exotic dark energy
\cite{Beng09,Lind10,Li11}. However one should remain cautious when
selecting the form of function $f(T)$. It is a well established fact
in our every day experience that weak-field gravitational bodies
like the sun or the earth do not experience violent instabilities
resulting in dramatic changes in their gravitational fields. So any
theory which results in such instabilities should be clearly ruled
out. It has been shown that some prototypes of $f(R)$ theory suffer
from these instabilities \cite{Dolg03} and a general condition for
the stability of such theories has been derived \cite{Faraoni06}.
Similarly one should consider stability of the theory in the
weak-field limit of $f(T)$ gravity. In this paper we show that
matter is generally stable in the context of modified teleparallel
gravity.

\section{Field equations}
In teleparallel gravity we need to define four orthogonal vector
fields named tetrad which form the basis of spacetime. The manifold
and the Minkowski metrics are related as
\begin{equation}
g_{\mu\nu}=\eta_{ij}e_{\mu}^{i}e_{\nu}^{j}
\end{equation}
where the Greek indices run from $0$ to $3$ in coordinate basis of
the manifold, the Latin indices run the same in tangent space of the
manifold and $\eta_{ij}=diag(+1,\,-1,\,-1,\,-1)$.  The connection in
teleparallel theory, the Weitzenbock connection, is defined as
\begin{equation}
\Gamma^{\rho}_{\,\,\,\,\mu\nu}=e^{\rho}_{i}\partial_{\nu}e^{i}_{\mu}
\end{equation}
which gives the spacetime a nonzero torsion but zero curvature in
contrast with general relativity. By this definition the torsion
tensor and its permutations are \cite{Haya79}
\begin{equation}
T^{\rho}_{\,\,\,\,\mu\nu}\equiv
e_{i}^{\rho}(\partial_{\mu}e_{\nu}^{i}-\partial_{\nu}e_{\mu}^{i})
\end{equation}
\begin{equation}
K^{\mu\nu}_{\quad\rho}=-\frac{1}{2}(T^{\mu\nu}_{\quad\rho}
-T^{\nu\mu}_{\quad\rho}-T_{\rho}^{\,\,\,\,\mu\nu})
\end{equation}
\begin{equation}
S^{\,\,\,\,\mu\nu}_{\rho}=\frac{1}{2}(K^{\mu\nu}_{\quad\rho}
+\delta^{\mu}_{\rho}T^{\alpha\nu}_{\quad\alpha}-\delta^{\nu}_{\rho}
T^{\alpha\mu}_{\quad\alpha}).
\end{equation}
Where $S^{\,\,\,\,\mu\nu}_{\rho}$ is called the superpotential. In
correspondence with Ricci scalar we define a torsion scalar as
\begin{equation}
T=S^{\,\,\,\,\mu\nu}_{\rho}T^{\rho}_{\,\,\,\,\mu\nu}
\end{equation}
so the gravitational action is
\begin{equation}
I=\frac{1}{16\pi G}\int d^{4}x\,|e|\, T
\end{equation}
where $|e|$ is the determinant of the vierbein $e^{a}_{\mu}$ which
is equal to $\sqrt{-g}$. Variation of the above action with respect
to the vierbeins will give the teleparallel field equations
\begin{equation}
e^{-1}\partial_{\mu}(ee_{i}^{\rho}S^{\,\,\,\,\mu\nu}_{\rho})
-e_{i}^{\lambda}T_{\,\,\,\,\mu\lambda}^{\rho}S^{\,\,\,\,\nu\mu}_{\rho}
+\frac{1}{4}e_{i}^{\nu}T=4\pi G
e_{i}^{\rho}\Theta^{\,\,\nu}_{\rho}
\end{equation}
Now similar to modifying the action of general relativity which $R$
is replaced by a general function $f(R)$, one can replace the
teleparallel action $T$ by a function $f(T)$. Doing this, the
resulting modified field equations are
$$e^{-1}\partial_{\mu}(eS^{\,\,\,\,\mu\nu}_{i})f^{\prime}(T)
-e^{\lambda}_{i}T^{\rho}_{\,\,\,\,\mu\lambda}S_{\rho}^{\,\,\,\,\nu\mu}f^{\prime}(T)
$$
\begin{equation}
+S_{i}^{\,\,\,\,\mu\nu}\partial_{\mu}(T)f^{\prime\prime}(T)
+\frac{1}{4}e^{\nu}_{i}f(T)=4\pi
G\,e^{\rho}_{i}\,\Theta^{\,\,\nu}_{\rho}
\end{equation}
where $\Theta^{\,\nu}_{\rho}$ is the energy momentum tensor of
matter. In what follows we set $4\pi G=1$.

\section{Matter Stability}
The main  motivation for modifying gravity in both teleparallel and
general relativity is the explanation of present time accelerated
expansion of the universe. If one considers a flat, homogeneous
Friedmann-Robertson-Walker universe, then the tetrads are
\begin{equation}
e^{i}_{\mu}=diag\Big(1,a(t),a(t),a(t)\Big)
\end{equation}
and the torsion scalar will be
\begin{equation}
T=-6\frac{\dot{a}^{2}}{a^{2}}=-6H^{2}
\end{equation}
From the field equation (9) one can derive the modified Friedmann
equation as \cite{Beng09}
\begin{equation}
12H^{2}f'(T)+f(T)=4\rho
\end{equation}
To achieve the present time acceleration, any added term to the
torsion scalar should be dominant at late times but negligible at
early times. In ref. \cite{Beng09} the form
$f(T)=T-\epsilon/(-T)^{n}$ has been proposed. This gives the correct
cosmological dynamics at late times without resorting to dark
energy.

Now we turn to the problem of matter stability. Following the above
discussion we promote the torsion scalar, $T$ to a general function
in the form
\begin{equation}
f(T)=T+\epsilon \varphi(T)
\end{equation}
where the parameter $\epsilon$ should be small to agree with recent
observational constraints. To study the matter stability of a model
of modified teleparallel gravity in the form (13), we begin by
taking the trace of field equation (9)

$$e^{-1}\partial_{\mu}(eS^{\,\,\,\,\mu\nu}_{\nu})f^{\prime}(T)
+S^{\,\,\,\,\mu\nu}_{\rho}\partial_{\mu}(e_{i}^{\rho})f'(T)e_{\nu}^{i}$$
\begin{equation}
+T f'(T) +S^{\,\,\,\,\mu\nu}_{\nu}
\partial_{\mu}(T)f^{\prime\prime}(T)+f(T)=\Theta
\end{equation}
Substituting (13), gives

$$e^{-1}\partial_{\mu}(eS^{\,\,\,\,\mu\nu}_{\nu})
(1+\epsilon\varphi')+S^{\,\,\,\,\mu\nu}_{\rho}\partial_{\mu}(e_{i}^{\rho})(1+\epsilon\varphi')e_{\nu}^{i}$$
$$+T(1+\epsilon\varphi')$$
 \begin{equation}
 +S^{\,\,\,\,\mu\nu}_{\nu}
\partial_{\mu}(T)\epsilon\varphi''+\Big(T+\epsilon
\varphi(T)\Big)=\Theta
\end{equation}
Note that Eqs. (14) and (15) correspond to the trace of the equation
of motion since the only kind of perturbations we take into account
are the ones of conformal factor \emph{i.e.} scalar modes.  Now we
apply this equation to the gravitational field of a weak-field
object like the sun or the earth. For such gravitational bodies, the
torsion scalar in \emph{linear perturbation} can be approximated by
\cite{Dolg03}
\begin{equation}
-T=\Theta+2\nabla_{\mu}T_{\nu}^{\,\,\,\mu\nu} +T_{1}
\end{equation}
where $T_{1}$ is the linear perturbation and $\nabla_{\mu}$ is the
covariant derivative with the Levi-Civita connection. This equation
followed from the fact that the torsion scalar $T$ and the Ricci
scalar $R$ only differ by a total divergence,
$R=-T-2\nabla_{\mu}T_{\nu}^{\,\,\,\mu\nu}$. The minus sign in (16)
comes from the fact that the torsion scalar is negative for a
homogeneous and isotropic weak field gravitational body.

The metric is also approximately can be taken as the Minkowski
metric plus some small perturbations
\begin{equation}
g_{\mu\nu}=\eta_{\mu\nu}+h_{\mu\nu}
\end{equation}
where we assume perturbations to be homogeneous and isotropic. Eq.
(17) means that the vierbeins can also locally be written in the
form
\begin{equation}
e^{\,\,i}_{\nu}=\delta^{\,\,i}_{\nu}+\tilde{e}_{\nu}^{\,\,i}
\end{equation}
where $\tilde{e}_{\nu}^{\,\,i}$ is a small perturbation in relation
to the trivial tetrad. We can describe the deviation from the flat
spacetime by [4]
\begin{equation}
\tilde{e}_{\nu}^{\,\,i}=\alpha\tilde{e}_{(1)\nu}^{\,\,i}+\alpha^{2}\tilde{e}_{(2)\nu}^{\,\,i}+...
\end{equation}
where $\alpha$ is a dimensionless parameter which labels the order
of perturbations. Inserting this expansion in (17), the
corresponding expansion of the metric is
\begin{equation}
g_{\mu\nu}=\eta_{\mu\nu}+\alpha(\tilde{e}_{(1)\mu\nu}+\tilde{e}_{(1)\nu\mu})+...
\end{equation}
and we have
$\tilde{e}_{(1)\nu}^{\,\,\rho}=\delta_{i}^{\rho}\tilde{e}_{(1)\nu}^{\,\,i}$
and
$\tilde{e}_{(1)\mu\nu}=\eta_{\mu\rho}\tilde{e}_{(1)\nu}^{\,\,\rho}$.\\
Here we consider only the first order or linear perturbations so
from now on we drop the subscript $(1)$ from the equations.\\
By perturbing the torsion in the form of equation (16) we'll have
the following equation in linear perturbation theory for the nearly
flat region inside a weak field celestial body (see Appendix for
proof)
\begin{equation}
\partial_{\mu}S^{\,\,\,\,\rho\mu}_{\rho}=A\Bigg(\frac{\dot{\Theta}
+\dot{T}_{1}}{2\sqrt{\Theta}}
-\frac{T_{1}\dot{\Theta}}{4\Theta\sqrt{\Theta}}+\sqrt{\Theta}\,\,\partial_{t}^{3}\,\tilde{e}_{\nu}^{\nu}+\frac{\dot{\Theta}}
{2\sqrt{\Theta}}\partial_{t}^{2}\,\tilde{e}_{\nu}^{\nu}\Bigg)
\end{equation}

where $A=\frac{3}{2\sqrt{6}}$ is a positive constant. Inserting (16)
, (18) and (21) in (15) and keeping only the terms linear in
perturbations, we get

$$\Bigg[\frac{A}{2\sqrt{\Theta}}+\frac{A\epsilon\varphi'}{2\sqrt{\Theta}}
+A\epsilon\varphi''\sqrt{\Theta}\Bigg]\dot{T_{1}}$$
$$+\Bigg[-A\epsilon(\varphi'-\varphi'')\Big(\frac{\dot{\Theta}
}{4\Theta\sqrt{\Theta}}\Big) +\epsilon\varphi'+2\Bigg]T_{1}$$
$$=-A(1+\epsilon\varphi')\Big(\frac{\dot{\Theta}}{4\Theta\sqrt{\Theta}}\Big)
-A(1+\epsilon\varphi')\Big[\partial_{t}(\tilde{e})+\sqrt{\Theta}\,\,\partial_{t}^{3}\,\tilde{e}_{\nu}^{\nu}\Big]
+A(2+\epsilon\varphi')\Big[\frac{\dot{\Theta}}
{2\sqrt{\Theta}}\partial_{t}^{2}\,\tilde{e}_{\nu}^{\nu}\Big]
$$
$$-A(1+\epsilon\varphi')\sqrt{\Theta}\partial_{t}(\tilde{e}_{i}^{\nu})
\delta_{\nu}^{i}+\frac{A}{2}\sqrt{\Theta}\epsilon\varphi''
\dot{\Theta}$$
\begin{equation}
-\frac{1}{2}\Theta\epsilon\varphi'-\frac{1}{4}\epsilon \varphi
\end{equation}
where $A=\frac{3}{2\sqrt{6}}$ and a dot denotes differentiation with
respect to time. Note that the perturbation equation in modified
teleparallel gravity, equation (22) is a first order differential
equation in contrast to the second order equations that appear in
$f(R)$ theories \cite{Faraoni06}. The right hand side of (22) is a
source term involving the matter content and also deviation from the
flat background as in (17) and (18). Equation (22) can be rewritten
in a concise form as
\begin{equation}
m\dot{T_{1}}+nT_{1}=\Pi
\end{equation}
where we have defined
 $$m\equiv\frac{A}{2\sqrt{\Theta}}+\frac{A\epsilon\varphi'}{2\sqrt{\Theta}}+A\epsilon\varphi''\sqrt{\Theta}$$
 $$n\equiv-A\epsilon(\varphi'-\varphi'')\Big(\frac{\dot{\Theta}}{4\Theta\sqrt{\Theta}}\Big)
+\epsilon\varphi'+2$$
$$\Pi\equiv-A(1+\epsilon\varphi')\Big(\frac{\dot{\Theta}}{4\Theta\sqrt{\Theta}}\Big)
-A(1+\epsilon\varphi')\Big[\partial_{t}(\tilde{e})+\sqrt{\Theta}\,\,\partial_{t}^{3}\,\tilde{e}_{\nu}^{\nu}\Big]
$$
$$+A(2+\epsilon\varphi')\Big[\frac{\dot{\Theta}}
{2\sqrt{\Theta}}\partial_{t}^{2}\,\tilde{e}_{\nu}^{\nu}\Big]-A(1+\epsilon\varphi')\sqrt{\Theta}\partial_{t}(\tilde{e}_{i}^{\nu})
\delta_{\nu}^{i}$$
\begin{equation}
\frac{A}{2}\sqrt{\Theta}\epsilon\varphi''
+\dot{\Theta}-\frac{1}{2}\Theta\epsilon\varphi'-\frac{1}{4}\epsilon
\varphi
\end{equation}
Let us make a comparison between values of the terms in $m$.  For a
typical gravitational body the energy-momentum scalar, $\Theta$, is
proportional to the mass density of the body and is positive
\cite{Dolg03}
\begin{equation}
\Theta\sim (10^{3}sec)^{-2}\Big(\frac{\rho_{m}}{g\, \,cm^{-3}}\Big)
\end{equation}
where $\rho_{m}$ is the mass density of the body. For example we
have $\rho_{m}=5.52\, g/cm^{3}$ for the earth and $\rho_{m}=1.41
\,g/cm^{3}$ for the sun. The value of $\epsilon$ is fixed in such a
way that it gives the correct cosmological dynamics at late times,
so it should be extremely small. For example, a common class of
functions that are popular in $f(T)$ literature is
\begin{equation}
f(T)=T-\frac{\mu^{2(n+1)}}{(T)^{n}}
\end{equation}
where $n$ is some real number and the $\mu$ parameter will be fixed
to a value that the model can reproduce the late time accelerated
expansion of the universe \cite{Beng09,Li11}. For this model we have
\begin{equation}
\mu^{-1}\sim 10^{18}\,sec
\end{equation}

From this it is obvious that the first term in $m$ is much larger
than the other two terms and we can safely neglect the second and
third terms. Doing this, equation (22) becomes

$$\dot{T_{1}}
+\Bigg[-\epsilon(\varphi'-\varphi'')\Big(\frac{\dot{\Theta}
}{2\Theta}\Big)+\frac{2\epsilon\varphi'\sqrt{\Theta}}
{A}+\frac{4\sqrt{\Theta}}{A}\Bigg]T_{1}$$
\begin{equation}
=\Big(\frac{2\sqrt{\Theta}}{A}\Big)\Pi
\end{equation}

 Let's consider the time evolution of perturbations. From the form of
differential equation (28) it is obvious that first order
perturbations, $T_{1}$ will grow with time if the coefficient of
$T_{1}$ in (28) is negative and decreases with time if the
coefficient is positive. Growing of perturbations with time will
mean that the torsion will rise very quickly and leads to strong
instability while a decreasing perturbations will mean that the
gravitational field will bounce back to its equilibrium state and so
the body is stable. The coefficient of $T_{1}$ in (28) is dominated
by the last term $\frac{4\sqrt{\Theta}}{A}$ due to extremely small
value of $\epsilon$. Note that $A$ and $\Theta$ are positive so from
this discussion it is obvious that the coefficient of $T_{1}$ will
always remain positive and as a result the matter in these types of
theories is always stable.

Now we turn our attention to the case of a radiation fluid. For this
type of matter the trace of the energy-momentum tensor, $\Theta$ is
vanishing. From (16) we have
$-T=\nabla_{\mu}T_{\nu}^{\,\,\,\mu\nu}+T_{1}=T'_{1}$ . Inserting
this in the trace of field equation (9) yields
\begin{equation}
\dot{T'_{1}}+p\,T'^{3/2}_{1}-q\,T'^{1/2}_{1}=0
\end{equation}
where by definition
$$p\equiv\Bigg(\frac{4+2\epsilon\varphi}
{A(1+\epsilon\varphi')}\Bigg)$$
\begin{equation}
q\equiv\Bigg(\frac{2\epsilon\varphi}{1+\epsilon\varphi'}\Bigg)
\end{equation}
Solving equation (29) for the time evolution of $T_{1}$ gives
\begin{equation}
T'_{1} \left( t \right) =\frac{q}{p}\tanh \Big(\frac{1}{2}\,t\sqrt
{p\,q}+\frac{C}{2}\sqrt {p\,q} \Big)^{2}
\end{equation}
which of course is always stable because the perturbations will
become constant after some time. Here $C$ is an integration
constant. The limiting value is given by
$q/p=A\epsilon\varphi/(2+\epsilon\varphi)$ which is extremely small
because of the value of $\epsilon$. Figure (1) shows the qualitative
behaviour of $T_{1}$ as given by equation (31).
\begin{figure}[htp]
\begin{center}\includegraphics{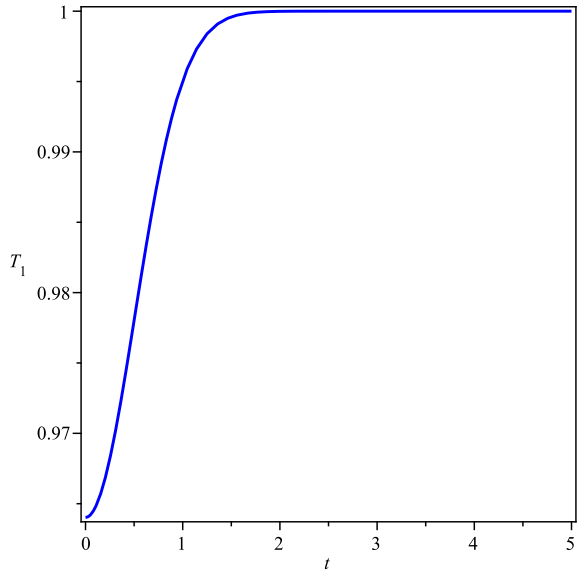} \vspace{6.5cm}
\end{center}
 \caption{\small {Qualitative behavior of the first order torsion scalar perturbation versus time for
  radiation matter with vanishing energy-momentum scalar.
   $T'_{1}$ will reach a constant value given by $q/p$
  and so the matter in this scenario is always stable.   }}
\end{figure}
\section{Conclusion}
From a geometric point of view, modifying gravity seems a necessary
task in order to explain recent positively accelerated expansion of
the universe. Any such modified theory, whether it is in the context
of general relativity or in teleparallel gravity,  may be expected
to show some strong deviation from the standard gravity at very high
energies and in strong-field regimes. This is because we still do
not have a proper theory of quantum gravity to describe the behavior
of gravitational interactions at those energies. On the other hand
any strong deviation from the standard gravity at low energies and
weak-field regimes immediately disqualify the theory because it will
contradicts well established weak-field experiments. One of these
experiments is the stability of weak-field celestial bodies or any
other weak gravity objects. In this paper we've investigated the
stability of such objects in the context of modified teleparallel
gravity. The analysis shows that there is no Dolgov-Kawasaki matter
instability in these type of theories. In contrast, in the
corresponding $f(R)$ theories a certain stability condition should
be met. This gives a great advantage to $f(T)$ theories over their
$f(R)$ counterparts because from matter stability viewpoint, there
is no limit on the form of functions that can be chosen to replace
the torsion scalar in the action of $f(T)$ theories. We note that we
have extended our analysis to the second order of perturbations and
we have observed that the matter is still stable in this scenario.\\

\appendix
\section{}
Here we present the proof of equation (23) for an almost flat region
inside a weak field gravitational body. For such an object the
tetrad and metric are given by equations (18) and (20) respectively.
Considering only the first order perturbations and dropping the
subscript we have the following equations for the torsion and
superpotential tensors
\begin{equation}
T_{\,\,\,\,\mu\nu}^{\rho}=\partial_{\mu}\tilde{e}_{\nu}^{\,\,\rho}-\partial_{\nu}\tilde{e}_{\mu}^{\,\,\rho}
\end{equation}
\\and\\

$$S^{\,\,\,\,\rho\mu}_{\nu}=\partial^{\rho}\tilde{e}_{\nu}^{\,\,\mu}-\partial^{\mu}\tilde{e}_{\nu}^{\,\,\rho}
-\delta_{\nu}^{\mu}(\partial^{\rho}\tilde{e}_{\sigma}^{\,\,\sigma}-\partial_{\sigma}\tilde{e}^{\sigma\rho})$$
\begin{equation}
+\delta_{\nu}^{\rho}(\partial^{\mu}\tilde{e}_{\sigma}^{\,\,\sigma}-\partial_{\sigma}\tilde{e}^{\sigma\mu})
\end{equation}
the tensor $\tilde{e}_{\nu}^{\,\,\mu}$ is not necessarily symmetric
but it has been shown that the anti-symmetric part of it has no
physical significance in the field equations so we assume it to be
symmetric here[4]. Furthermore for an almost flat region inside a
star, we can safely assume that both the background and the first
order correction are homogeneous and isotropic. In that case the
torsion and its perturbation does not depend on spatial coordinates
and we have $\partial_{\mu}\rightarrow \partial_{t}$. Also for a
homogeneous and isotropic perturbation the first order correction of
the tetrad has the form
\begin{equation}
\tilde{e}_{\nu}^{\,\,\mu}=diag(1,b,b,b)
\end{equation}
and $b$ only depends on time. Substituting this in (A1) and (A2), we
can find the torsion scalar as
\begin{equation}
T=S^{\rho\mu\nu}T_{\rho\mu\nu}=-6\dot{b}^{2}
\end{equation}
Up to the first order in perturbations, the second term in the R.H.S
of Eq. (16) will be $\nabla_{\mu}T_{\nu}^{\,\,\,\mu\nu}=3\ddot{b}$.
On the other hand the only non zero components of the superpotential
tensor are all the same (up to a sign) and proportional to
$\sqrt{T}$, in particular we have
\begin{equation}
\partial_{\mu}S_{\nu}^{\mu\nu}=\frac{3}{2}\ddot{b}
\end{equation}
so we will have the relation
\begin{equation}
\partial_{\mu}S_{\nu}^{\mu\nu}=\frac{3}{2\sqrt{6}}\partial_{t}{(\sqrt{-T})}
\end{equation}
substituting from (16), equation (23) is obtained.
\section*{Acknowledgement}
It is really our pleasure to thank Professor Valerio Faraoni for
fruitful discussion.\\


\begin{thebibliography}{11}
\bibitem{Ein30}
A. Einstein, Math. Annal. 102, 685 (1930). For an english
translation, see A. Unzicker and T. Case, [arXiv:physics/0503046v1].
\bibitem{Weitz23}
R. Weitzenbock, Invariance Theorie, Nordhoff, Groningen, 1923.
\bibitem{Haya79}
K. Hayashi and T. Shirafuji, Phys. Rev. D 19, 3524 (1979); K.
Hayashi and T. Shirafuji, Phys. Rev. D 24, 3312 (1981);
\bibitem{Ald10}
R.Aldrovandi and J. G. Pereira, An Introduction to Teleparallel
Gravity, Instituto de Fisica Teorica, UNSEP, Sao Paulo.
\bibitem{SN98} A. G. Riess et al. [Supernova Search Team
Collaboration], Astron. J. 116, 1009 (1998); S. Perlmutter et al.
[Supernova Cosmology Project Collaboration], Astrophys. J. 517, 565
(1999);
\bibitem{FR}
S. M. Carroll, V. Duvvuri, M. Trodden and M. S. Turner, Phys. Rev. D
70, 043528 (2004); S. Nojiri and S. D. Odintsov, Gen. Rel. Grav. 36,
1765 (2004); S. Nojiri and S. D. Odintsov, Phys. Rev. D 68, 123512
(2003); G. Allemandi, M. Capone, S. Capozziello and M. Francaviglia,
Gen. Rel. Grav. 38, 33 (2006); S. Nojiri and S.D. Odintsov,
[arXiv:hep-th/0601213]; T. P. Sotiriou, Class. Quant. Grav. 23, 5117
(2006); T. P. Sotiriou and S. Liberati, Annals Phys. 322, 935
(2007); T. P. Sotiriou and V. Faraoni, Rev. Mod. Phys. 82, 451
(2010); S. Tsujikawa, Lect. Notes Phys. 800, 99 (2010); S. Nojiri
and S. D. Odintsov, Phys. Rept. 505, 59 (2011).

\bibitem{FT}
R. Ferraro and F. Fiorini, Phys. Rev. D75, 084031 (2007); R. Ferraro
and F. Fiorini, Phys. Rev. D78, 124019 (2008); P. Wu and H. W. Yu,
Eur. Phys. J. C 71, 1552 (2011); R. Zheng and Q. G. Huang, JCAP
1103, 002 (2011); K. Bamba, C. Q. Geng, C. C. Lee and L. W. Luo,
JCAP 1101, 021 (2011); T. Wang, Phys. Rev. D84, 024042 (2011);  P.
Wu and H. W. Yu, Phys. Lett. B 693, 415 (2010); G. R. Bengochea,
Phys. Lett. B 695, 405 (2011); P. Wu and H. W. Yu, Phys. Lett. B692,
176 (2010); S. H. Chen, J. B. Dent, S. Dutta and E. N. Saridakis,
Phys. Rev. D 83, 023508 (2011); J. B. Dent, S. Dutta and E. N.
Saridakis, JCAP 1101, 009 (2011); X. C. Ao, X. Z. Li and P. Xi,
Phys. Lett. B694, 186 (2010); Y. Zhang, H. Li, Y. Gong and Z. H.
Zhu, JCAP 1107, 015 (2011); R. Ferraro and F. Fiorini, Phys. Lett.
B702, 75 (2011); H. Wei, X. P. Ma and H. Y. Qi, Phys. Lett. B703, 74
(2011); P. Wu and H. Yu, Phys. Lett. B703, 223 (2011); C. Q. Geng,
C. C. Lee, E. N. Saridakis, Y. P. Wu, Phys. Lett. B704, 384 (2011);
S. Capozziello, V. F. Cardone, H. Farajollahi and A. Ravanpak, Phys.
Rev. D 84, 043527 (2011); K. Karami and A. Abdolmaleki,  JCAP 04,
007 (2012); M. Li, R. -X. Miao, Y. -G. Miao, JHEP 1107, 108 (2011);
R. -X. Miao, M. Li and Y. -G. Miao, JCAP 11, 033 (2011).

\bibitem{Beng09}
G. R. Bengochea and R. Ferraro, Phys. Rev. D 79, 124019 (2009).
\bibitem{Lind10}
E. V. Linder, Phys. Rev. D 81, 127301 (2010).
\bibitem{Li11}
B. Li, T. P. Sotiriou and J. D. Barrow, Phys. Rev. D 83, 064035
(2011).
\bibitem{Dolg03}
A. D. Dolgov and M. Kawasaki, Phys. Lett. B 573, 1 (2003).
\bibitem{Faraoni06}
V. Faraoni, Phys. Rev. D 74 104017 (2006)

\end{thebibliography}
\end{document}